 \documentclass[prl,aps,twocolumn,showpacs]{revtex4}
\usepackage[dvips]{graphicx}
\usepackage{dcolumn}
\newcommand{\be}{\begin{equation}}
\newcommand{\ee}{\end{equation}}
\newcommand{\bea}{\begin{eqnarray}}
\newcommand{\eea}{\end{eqnarray}}
\newcommand{\nn}{ \nonumber}

\begin{document}
\topmargin=-25mm
%   \Large

\title{Inelastic electron transport in polymer nanofibers}

\author{Natalya A. Zimbovskaya }

\affiliation
{Department of Physics and Electronics, University of Puerto 
Rico-Humacao, CUH Station, Humacao, PR 00791;\\ and
Institute for Functional Nanomaterials, University of Puerto Rico, San Juan, 
PR 00931}
%%\\ and Institute for Nanoelectronics and
%%Computing, Purdue University, West Lafayette, IN 47907}

\begin{abstract}
 In this paper we present theoretical analysis of the  electron transport 
in conducting polymers being in a metal-like state. 
We concentrate on the study of the effects of 
temperature on  characteristics of the transport. We treat a conducting
polymer in the metal state as a network of metallic-like grains 
embedded in poorly conducting environment which consists of randomly 
distributed polymeric chains. We carry out the present studies assuming 
that the intergrain conduction is mostly provided by electron quantum 
tunneling via intermediate states localized on polymer chains
between the grains. To analyze the effects of temperature on this kind of 
electron intergrain transport we represent the thermal environment as a phonon
bath coupled to the intermediate state. The electron transmission is computed
using the Buttiker model within the scattering matrix formalism. This approach
is further developed, and the dephasing parameter is expressed in terms of
relevant energies including the thermal energy. It is shown that temperature
dependencies of both current and conductance associated with the above
transport mechanism differ from those typical for other conduction mechanisms 
in conducting polymers. This could be useful to separate out the contribution
from the intergrain electron tunneling to the net electric current in
transport experiments on various polymer nanofibers. The proposed model could
be used to analyze inelastic electron transport through molecular junctions.
   \end{abstract}

%%{71.18.+y, 71.20-b, 72.55+s}

\date{\today}
\maketitle

\section{i. introduction}

At present, electronic transport properties of conducting polymers such as 
doped polyacetylen and polyaniline-polyethylene oxides attract an intense 
interest and attention of the research community \cite{1,2}. These materials, 
as well as carbon nanotubes, are expected to have various applications in 
fabrication of nanodevices. Significant efforts have been applied to study 
conduction mechanisms in the polymers. This is a rather complex topic
for electron transport in conducting polymers shows both metallic and 
non-metallic features, and various transport mechanisms contribute to the
resulting pattern.

An important contribution to
the conduction in these substances is provided by the phonon assisted electron 
hopping between the conducting islands or variable range hopping between localized
electronic states \cite{2,3}. The effect of these transport mechanisms strongly
depends on the intensity of stochastic nuclear motions. The latter increases as 
temperature rises, and this brings a significant enhancement of the corresponding
contribution to the conductivity. The temperature dependence of the ``hopping" 
conductivity $\sigma (T) $ is given by the Mott's expression \cite{4}:
  \be %f1
 \sigma (T) = \sigma (0) \exp \left[-(T_0/T)^p \right]   \label{1}
  \ee
  where $T_0$ is the characteristic temperature of a particular material, and
the parameter $p$ takes on values $0.25,\,0.33 $ or $0.5 $ depending on the
dimensions of the hopping processes.  
Also, it was proposed that phonon-assisted transport in low-dimensional structures 
such as nanofibers and nanotubes may be substantially influenced due to electron
interactions \cite{5,6}. This results in the power-low temperature dependencies
of the conductance $G(T)$ at low values of the bias voltage $V \ (eV < kT,\ k $
being the Boltzmann constant), namely: $G \sim T^\alpha. $ Experimental
data for the conductance of some nanofibers and nanotubes match this power-low
reasonably well, bearing in mind that the value of the exponent $ \alpha$ varies 
within a broad range. For instance, $\alpha $ was reported to accept values 
about $0.35 $ for carbon nanotubes \cite{7}, and $\alpha \sim 2.2 \div 7.2 $ for 
various polyacetylene nanofibers \cite{8,9}.

In general, hopping transport mechanism is very important in disordered 
materials with localized states. For this kind of transport phonons play  part
of a source of electrical conductivity. Accordingly, the hopping contribution
to the conductivity always increases as temperature rises, and more available
phonons appear. When polymers are in the insulating state, the hopping transport
mechanism predominates and determines the temperature dependencies of transport
characteristics.

In metallic state of conducting polymers free charge carriers appear, and their
motion strongly contributes to the conductance. While moving, the charge carriers 
undergo scattering by phonons and impurities. This results in the conductivity
stepping down. Metallic-like features in the temperature dependencies of dc
conductivity of some polymeric materials and carbon nanotubes were repeatedly
reported. For instance, the decrease in the conductivity upon heating was observed
in polyaniline nanofibers Refs. \cite{10,11} and carbon nanotubes \cite{12}. 
 However, this electron diffusion is not the only transport
mechanism in the metallic state.
The conducting polymers could be described 
as granular metals, where conducting metallic-like islands (grains) made out 
of densely packed polymer chains are embedded in the amorphous poorly 
conducting environment where the chains are disorderly arranged \cite{1}. 
  Prigodin and Epstein suggested that the electron transport between metallic 
islands  mostly occurs as a result of electron resonance tunneling through 
intermediate states localized on the polymer chains in between the grains \cite{13}.
The effect of phonons on this kind of electron transport may be very significant. 
  These phonons bring an inelastic component to the intergrain 
current and underline the interplay between the transport by the electron 
tunneling and the thermally assisted dissipative transport. Also, they may 
cause some other effects, as was shown while developing the theory of 
conduction through the molecules \cite{14,15,16,17}. 

Here, we concentrate on the analysis of temperature dependencies of the electric
current and conductance associated with the resonance tunneling transport 
mechanism. The analysis is motivated by an assumption that various conduction
mechanisms may simultaneously contribute to the charge transport in conducting
polymers, and their relative effects could significantly differ depending on
the specifics of synthesis and processing of polymeric materials. The temperature
dependencies of the resulting transport characteristics may help to identify
the predominating transport mechanism for a paricular sample under particular
conditions, providing a deeper insight in the nature of electron transport in
conducting polymers.

The issue is of particular  importance  because the relevant transport 
experiments are often implemented at room temperature (see e.g. \cite{18}), 
so that the influence of phonons cannot be disregarded.
Therefore we study the effect of temperature (stochastic nuclear motions)
on the  resonance electron tunneling between metalliclike grains (islands)
in polymer nanofibers.
We apply the obtained results to analyze electron transport in conducting
polymers. However, these results could be easily adapted to study some phonon
induced effects in the electron transport through metal-molecule junctions and
other kinds of quantum dots coupled to the source and drain reservoirs.

\section{ii. model and results}
 
The  transmission coefficient for the electron intergrain
resonance tunneling is determined with
the probability of finding the resonance state. The latter is estimated as 
$ P \sim \exp(-L/\xi)$  $(L $ is the average distance between the
adjacent grains, and $ \xi $ is the localization length for electrons),
and takes values much smaller than unity but much greater than the 
transmission probability for sequental hoppings along the chains, $ P_h 
\sim \exp (-2L/\xi) $ \cite{19}. The probability for existence of a resonance 
state at a certain chain is rather low, so only a few out of the whole set
of the chains connecting two grains are participating in the process of 
intergrain electron transport. Therefore one could assume that any two metallic
domains are connected by a single chain providing an intermediate state for
the resonance tunneling. All remaining chains can be neglected for they
poorly contribute to the transport compared to the resonance chain.
Within this approach the intergrain conduction strongly resembles the electron
 conduction through a molecular bridge connecting two metallic leads, and 
similar formalism could be employed to compute it.

Correspondingly, in calculations of the current 
we employ the formula which describes the electronic transport through a junction
including two leads (adjacent grains) and the intermediate state coupled to them.
We treat the grains as free electron reservoirs in thermal equilibrium. This
assumption is justified when the intermediate state (the bridge) is weakly
coupled to the leads and conduction is much smaller than the  quantum 
conductance $G_0 = 2e^2/h\ (e,h $ are the electron charge, and the Planck
constant, respectively). Due to the low probabilities for the resonance
tunneling between the metallic islands in the conducting polymers the above 
assumption may be considered as a reasonable one. So, we can employ the 
well-known expression for the electron current through the junction \cite{20},
and we write:
  \be % f2,1,18
 I = \frac{2en}{h} \int_{-\infty}^\infty dE T(E) [f_1(E) - f_2(E)].   \label{2}
  \ee  
  Here, $ n $ is the number of the working channels in the fiber, $f_{1,2} (E)$ 
are Fermi functions taken with the different contact chemical potentials 
$ \mu_{1,2} $ for the grains. The chemical potentials differ due to the  
bias voltage $ \Delta V $ applied across the grains:
  \be % f3,2,19
 \mu_1 = E_F + (1 - \eta) e \Delta V; \qquad
 \mu_2 = E_F - \eta e \Delta V.                     \label{3}
  \ee
   The parameter $ \eta $ characterizes how the voltage $ \Delta V $ is divided 
between the grains; $ E_F $ is the equilibrium Fermi energy of the system 
including the pair of grains and the resonance chain in between, and $ T (E) $ 
is the electron transmission function.

Realistic polymer nanofibers have diameters within the range $ 20\div 100$nm,
and lengths of the order of a few microns. This is much greater than the 
typical size of both metallic-like grains and intergrain separations which 
take on values $ \sim 5\div 10$nm (see e.g.  \cite{18,21}). Therefore, we may 
treat a nanofiber as a set of working channels connected in parallel, any 
single channel being a sequence of grains connected with the resonance polymer 
chains. The net current in the fiber is the sum of currents flowing in these 
channels, and the voltage $ V$ applied across the whole fiber is distributed 
among sequental pairs of grains along a single channel. So, the voltage 
$ \Delta V$ applied across two adjacent grains could be roughly estimated as 
$ \Delta V \sim V{L/L_0} $ where $ L $ is the average separation between the 
grains, and $ L_0 $ is the fiber length. 
In realistic fibers the ratio $ \Delta V/V $ may take on values of 
the order of $  10^{-2} \div 10^{-3} . $ 

To proceed we must compute $T(E) $. An important issue in calculation of
the transmission is the effect of stochastic nuclear motions in the
environment of the resonance state. 
  When the dissipation is strong (e.g. within the strong thermal coupling limit),
the inelastic (hopping) 
contribution to the intergrain current predominates, replacing the
coherent tunneling dominating at weak dephasing. Typically, at room 
temperatures the intergrain electron transport in conducting polymers occurs
within an intermediate regime, when both coherent and incoherent contributions
to the electron transmission are manifested. 

The general approach to the electron transport  studies 
in the presence of dissipation is the reduced dynamics density-matrix
formalism (see, e.g., Refs. \cite{22} and \cite{23}). This microscopic
computational approach has the advantages of being capable of 
providing the detailed 
dynamics information. However, this information is usually more redundant
than necessary, as far as standard transport experiments in conducting
polymer nanofibers are concerned. There exists  an
alternative approach using the scattering-matrix formalism and the 
phenomenological Buttiker dephasing model \cite{24,25}. Adopting this
phenomenological model we are able to analytically treat the problem, and
the results agree with those obtained by
means of more sophisticated computational methods, as was demonstrated in 
the earlier work \cite{14}. 
 However, this alternative approach has some significant shortcomings. Its 
main disadvantage is that the dissipative effects are described in terms
of a phenomenological dephasing parameter $\epsilon $ whose dependence of
the characteristic factors affecting the transport (such as the
temperature, the electron-phonon coupling strength
and some others) remains unclear. 
Here, we carry out our analysis within the framework of the phenomenological
Buttiker's dephasing model but we modify the latter to elucidate the relation
of the dephasing parameter $\epsilon $ to the relevant energies characterizing
the electron transport in the considered system.

In studies of the intergrain electron transport in conducting polymers the
 ``bridge" between two grains inserts a single 
electron state. Therefore we may treat the electron transport as a combination
of tunneling through two barriers (the first one separates the left metallic
domain from the intermediate state in the middle of the resonance chain, and
the second separates this state from the right grain, supposing the transport
is from the left to the right) affected by inelastic scattering at the bridge,
as shown in the Fig. 1. The barriers are represented by the squares, and the 
triangle in between imitates a scatterer coupling the bridge to a dissipative 
electron reservoir.

An electron could be injected into this system, and/or leave from there via
four channels indicated in this Figure. Incoming particle fluxes 
$(J_i)$ are related to those outgoing from the system $(J_j')$ by means of the
transmission matrix $T, $ \cite{24,25}
  \be %f3
  J_j' = \sum_i T_{ji} J_i, \qquad 1\leq i,  j \leq 4.     \label{4}
  \ee
  Off-diagonal matrix elements $T_{ji} (E) $ are probabilities for the electron
to be transmitted from the channel $ i $ to the channel $j, $ whereas diagonal
matrix elements $T_{ii}(E) $ are probabilities for its reflection back to the
channel $i $. To provide charge conservation, the net particle flux
in the channels connecting the system with the reservoir must be zero. So we
have:
  \be %f5,4
J_3 + J_4 - J_3' - J_4' = 0.            \label{5}
  \ee 
 The transmission function $ T(E) $ relates the particle flux outgoing from 
the channel $2$ to the flux incoming to the channel $1,$ namely:
  \be %f6,5
J_2' = T(E) J_1.                \label{6}
 \ee
   Using Eqs. (4) and (5) we can express the transmission function in
terms of the matrix elements of the scattering matrix $S$ relating the 
outgoing wave amplitudes $b_1',b_2',a_3', a_4'$ to the incident ones 
$b_1,b_2,a_3, a_4:\ T_{ij} = |S_{ij}|^2. $ In the considered case 
of a single site bridge the $S$ matrix takes the form \cite{14}:
  \be %f7,6
 S = Z^{-1}\left(
   \begin{array}{cccc}
r_1 + \alpha^2 r_2 & \alpha t_1 t_2 & \beta t_1 & \alpha\beta t_1 r_2 
   \\
\alpha  t_1 t_2 & r_2 +\alpha^2 r_1 & \alpha\beta r_1 t_2 &  \beta t_2
   \\
 \beta t_1 & \alpha \beta r_1 t_2 & \beta^2 r_1 & \alpha r_1 r_2 - \alpha
   \\
\alpha \beta t_1 r_2 & \beta t_2 & \alpha r_1 r_2 - \alpha & \beta^2 r_2
    \end{array} \right) .                   
                            \label{7}     \ee
  where $Z = 1 - \alpha^2 r_1 r_2,\ \alpha = \sqrt{1- \epsilon}, \ \beta =
\sqrt\epsilon,$ $r_{1,2}$ and $t_{1,2} $ are the amplitude transmission an
reflection coefficients for the barriers $(|t_{1,2}|^2 + |r_{1,2}|^2 = 1),$
and the parameter $ \epsilon $ characterizes the dephasing strength. This 
parameter takes values within the range $[0,1],$
so that $ \epsilon = 0$ corresponds to
the completely coherent and $ \epsilon = 1 $ to the fully incoherent transport.

\begin{figure}[t] % fig. 1
\begin{center}
\includegraphics[width=2.8cm,height=8cm,angle=-90]{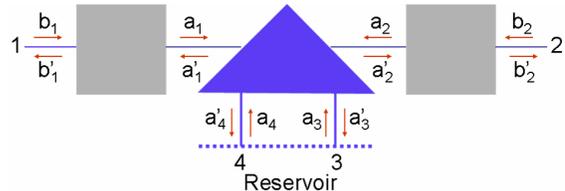}%{n20_5.eps}
\caption{ Schematic drawing illustrating the intergrain electron transport
in the presence of dissipation \cite{24}.
}  
\label{rateI}
\end{center}
\end{figure}

When the bridge is detached from the dephasing reservoir $ T(E) = |S_{12}|^2.$
On the other hand, in this case we can employ a simple analytical expression 
for the electron transmission function \cite{26,27,28}
  \be %f8,6,7
  T(E) = 4 \Delta_1(E) \Delta_2 (E) |G(E)|^2,      \label{8}
  \ee
 where $\Delta_{1,2} (E) = - \mbox{Im} \Sigma_{1,2} (E). $ In this expression,
self-energy terms $ \Sigma_{1,2} $ appear due to the coupling of the metallic
grains to the intermediate state (the bridge).
  The retarded Green's function for a single-site bridge could be approximation
as follows:
  \be %f9,8
  G(E) = \frac{1}{E -E_1 +i\Gamma}         \label{9}
  \ee
  where $E_1 $ is the site energy. The width of the resonance level between
the grains is described by the parameter $ \Gamma = \Delta_1 + \Delta_2 + 
\Gamma_{en}\ (\Gamma_{en}$ describes the effect of the environment).

Equating the expression (8) and $ t_1^2  t_2^2 $ we arrive at the following
expressions for the tunneling parameters $ \delta_{1,2} (E) :$
   \be %f10,9
 \delta_{1,2}(E)\equiv t_{1,2}^2 = 
\frac{2 \Delta_{1,2}}{\sqrt{(E- E_1)^2 + \Gamma^2}} . \label{10}
  \ee

Using this result we easily derive the general expression for the electron
transmission function:
  \be %f11,10,3
  T (E) = \frac{g(E)(1+ \alpha^2) [g(E)(1+ \alpha^2) + 1 
- \alpha^2]}{[g(E)(1 - \alpha^2) + 1 + \alpha^2]^2}         \label{11}
   \ee
  where:
    \be %f12,11,4
  g(E) = %2 \sqrt{\Delta_1\Delta_2\big/[(E - E_1)^2 + \Gamma^2]},
2 \sqrt{\frac{\Delta_1\Delta_2}{(E - E_1)^2 + \Gamma^2}},   \label{12}
  \ee

In studying the effect of the stochastic nuclear motions (effect of the
temperature) on the intergrain electron transport we may treat the former
as a phonon bath. In the present work we assume
that the intermediate site (the bridge) is directly coupled to the phonons 
produced by nuclear motions. Such approach was used before in some papers 
analyzing dissipative effects in the electron conduction through molecules 
(see e.g. \cite{15}), and we apply it here. Then $\Gamma_{en} $
equals $ \Gamma_{ph} $ where $ \Gamma_{ph} (E) $ is the imaginary part of
the electronic self-energy correction due to the electron-phonon interaction.
 However, in analyzing the electron transport in polymers, as well as in 
molecules, one must keep in mind that besides  the bridge sites there always 
exist other nearby sites. In some cases the presence of such sites may strongly 
influence the effects of the stochastic nuclear motions on the characteristics 
of the electron transport. This may happen when the nearby sites somehow 
``screen" the bridge sites from the direct interaction with the phonon bath 
\cite{17}. 
 We elucidate some effects which could appear in the electron transport in 
conducting polymer fibers in the case of such indirect coupling of the bridge 
state to the phonon bath elsewhere \cite{29}.

To achieve better understanding of the effect of temperature in the electron 
transport in polymer fibers within the adopted approach we must express the 
dephasing strength $\epsilon $ in terms of relevant energies. As shown before 
(see \cite{30}), $\epsilon$ could be written in the form:
  \be %f13,12,5
 \epsilon =\frac{ \Gamma_{ph}}{\Gamma}.    \label{13}
  \ee

In further calculations we assume that the phonon bath is characterized by the 
continuous spectral density $J (\omega)$ of the form \cite{31}:
  \be %f14,13,6,13
J(\omega) = J_0 \frac{\omega}{\omega_c} \exp \left(-\frac{\omega}{\omega_c}
\right )             \label{14}
  \ee
  where $ J_0 $ describes the electron-phonon coupling strength, and 
$ \omega_c $ is the cut-off frequency of the bath characterizing the thermal 
relaxation rate of the latter.

Starting from the corresponding result of the earlier works \cite{15,27,28} and 
using this expression \ref{14} we may present $ \Gamma_{ph} (E) $ as follows:
  \bea %f15,14,7
 \Gamma_{ph}& =& 2\pi J_0 \int d\omega 
\frac{\omega}{\omega_c} \exp \left(-\frac{\omega}{\omega_c}
\right )
   \nn\\ &\times&
\big \{N(\omega) [\rho_{el}(E-\hbar\omega)
 +\rho_{el}(E+\hbar\omega)]
 \nn\\& +&
 [1-n(E-\hbar\omega) ]\rho_{el} (E-\hbar\omega)
 \nn\\& +&
 n(E + \hbar\omega) \rho_{el} (E+\hbar\omega) \big\}.      \label{15}
 \eea
  Here,
 \be %f16,15,8
n(E) = \frac{\Delta_1 f_1(E) + \Delta_2 f_2 (E)}{\Delta_1 + \Delta_2}
             \label{16}     \ee
 $\rho_{el} (E) =( -1/\pi) \mbox{Im} (E - E_1 + i \Gamma)^{-1} $ is the 
electron density of states, $ N(\omega) $ is the Bose-Einstein distribution  
function for the phonons at the temperature $T. $ The asymptotic expression 
for the self-energy term $ \Gamma_{ph} $ depends on the relation between two 
characteristic energies, namely: $ \hbar \omega_c $ and $ kT \ (k$ is the 
Boltzmann constant). At moderately high temperatures $(T\sim 100\div 300K),$ 
which  are typical for the experiments on electrical characterization of polymer 
nanofibers $kT \sim 10\div 30 meV.$ This is significantly greater than typical 
values of $ \hbar \omega_c \ (\hbar\omega_c \sim 1 meV $ \cite{15}). Therefore 
in further calculations we assume $ \hbar \omega_c\ll kT. $ Under this 
assumption, the main contribution to the integral over $ \omega $ in the Eq. 
(15) originates from the region where $ \omega\ll \omega_c \ll kT/\hbar, $ 
and we can use the following approximation:
  \be % f17,16,9,14
 \Gamma_{ph} (E) = \frac{2 \Gamma \Lambda (J_0,\omega_c, T)}{(E- E_1)^2 + 
\Gamma^2}.          \label{17}
  \ee
  Here, 
   \be % f18,17,10,11,16
\Lambda = \frac{4 J_0}{\hbar \omega_c} (kT)^2 \zeta 
\left(2; \frac{kT}{\hbar\omega_c} + 1\right)           \label{18}
  \ee 
 where $\zeta (2; kT/\hbar\omega_c + 1) $ is the Riemann $ \zeta $ function: 
  \be %f19,18,11,12
\zeta = (2;kT/\hbar\omega_c + 1) = \sum_{n=1}^\infty \frac{1}{(n+ kT/\hbar
\omega_c)^2}.             \label{19}
  \ee
  Under $ \hbar\omega_c \ll kT, $ we may apply the estimation $ \Lambda 
\approx 4k TJ_0. $

Solving the equation (17) we obtain a reasonable asymptotic expression for 
$ \Gamma_{ph}: $ 
   \be %f20,19,12
\Gamma_{ph} = \frac{\Delta_1 + \Delta_2}{2} \frac{\rho^2(1 + \sqrt{1 + 
\rho^2})}{4\big(\frac{E - E_1}{\Delta_1 + \Delta_2}\big)^2 + (1 + \sqrt{1 + 
\rho^2})^2} .                \label{20}
  \ee
  where $ \rho^2 = 8 \Lambda/(\Delta_1 + \Delta_2)^2. $
 Substituting this expression into (13) we arrive at the result for the 
dephasing strength $ \epsilon :$
  \be %f21,20,13
 \epsilon = \frac{1}{2}\frac{\rho^2(1 + \sqrt{1 + \rho^2})}{4\big
(\frac{E 
- E_1}{\Delta_1 + \Delta_2}\big)^2 +\frac{1}{2} (1 + \sqrt{1 + \rho^2})^3} 
       \label{21} \ee
  This expression shows how the dephasing parameter depends on the temperature 
$ T ,$ the electron-phonon coupling strength $ J_0, $  and the energy $ E. $ In 
particular, it follows from the Eq. (21) that $ \epsilon $ reachs its maximum 
at $ E = E_1, $ and the peak value of this parameter is given by:
  \be %f22,14,12,13,17
\epsilon_{max} = \frac{\sqrt{1 + \rho^2}-1}{\sqrt{1 + \rho^2}+1}.  \label{22}
  \ee 
  The obtained result enables us to analyze the temperature dependencies of the
electric current and conductance of the doped polymer fibers assuming that the
resonance tunneling predominates in the intergrain electron transport in the 
absence of phonons. 

\section{iii. discussion}

 The maximum value of the dephasing strength is determined with two parameters, 
namely, $ T $ and $ J_0. $ As illustrated in the Fig. 2,  
$ \epsilon_{\max}$ increases when the temperature rises, and it takes on 
greater values when the electron-phonon interaction is getting stronger. 
This result has a clear physical sense. Also, as follows from the
Eq. \ref{21},
the dephasing parameter  exhibit a peak at $ E = E_1 $ whose shape 
is determined by the product $ kTJ_0 .$ When either $ J_0$ or $ T $ or both 
enhance, the peak becomes higher and its width increases. 
The manifested energy dependence of the dephasing strength allows 
us to resolve a difficulty occuring when the inelastic contribution to the 
electron transmission function is estimated using the simplified 
approximation of the parameter $ \epsilon $ as a constant. 
Within such approximation, a significant 
rate of phase randomization appears in the electron transport between the 
metallic-like islands when $ \epsilon $ reachs values $ \sim 0.3\div0.5 $ or 
greater \cite{30}. Then the peak at the electron transmission is eroded, and 
current-voltage characteristics become linear. 
 Therefore, to keep in consideration a distinguishable coherent contribution 
to the current, one must assume $ \epsilon $ to take on small values. 
For instance, the agreement with the experimental data reported in the work 
\cite{18} was achieved assuming $ \epsilon = 0.05 $ \cite{32}. When the 
experiments are carried out at room temperatures such small values of 
$ \epsilon $ imply very weak electron-phonon coupling strength. This 
implication could hardly be given a reasonable physical explanation.
On the contrary, if the energy dependence of $\epsilon$ is accounted for, 
the peak in the electron transmission at $E=E_1$ may be still distinguishable
at moderately high values of the electron-phonon coupling strength $J_0.$

\begin{figure}[t]
\begin{center}
\includegraphics[width=5cm,height=9.2cm,angle=-90]{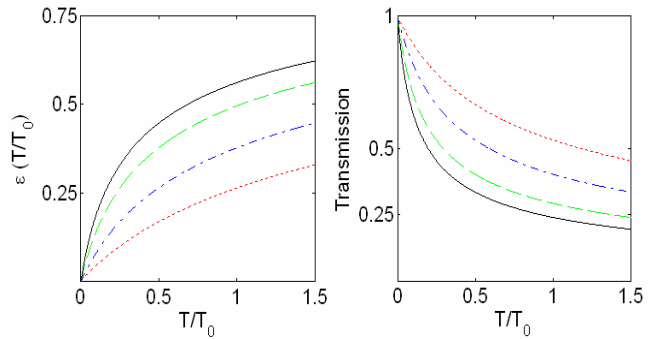}%%{n21_1.eps}
\caption{ Temperature  dependencies of the the dephasing parametr $\epsilon$ 
(left panel) and the electron transmission $T $ (right panel).
 The curves are plotted at $ T_0 = 50K,\ \Delta_1 = \Delta_2 = 4 meV,\ 
E = E_1 = 0,\  J_0 = 9.0 meV$ (solid line), $6.0 meV $ (dashed line), $3 meV $ 
(dash-dotted line), $1.5  meV$ (dotted line). 
}  
\label{rateI}
\end{center}
\end{figure}

Current-voltage characteristics and voltage dependencies of the conductance
$ G = dI/dV $ computed using the expressions \ref{2}, \ref{11}, \ref{21} are 
presented in the Fig. 3. We see that as the electron-phonon coupling 
strengthens, the I-V curves lose their specific shape typical for the coherent
tunneling through the intermediate state. They become closer to
straight lines corresponding to the Ohmic law. At the same time the maximum in 
the conductance originating from the intergrain tunneling gets eroded due to the
effect of phonons. These are the obvious results discussed in some earlier
works (see e.g. \cite{14}). The relative strength of the electron-phonon
interaction is determined by the ratio of the electron-phonon coupling constant
$J_0 $ and the self-energy terms describing the coupling of the intermediate 
state (bridge) to the leads $\Delta_{1,2}.$ The effect of phonons on the
electron transport becomes significant when $J_0> \Delta_{1,2}.$

\begin{figure}[t]  %%%fig. 3
\begin{center}
\includegraphics[width=9.8cm,height=9.2cm,angle=-90]{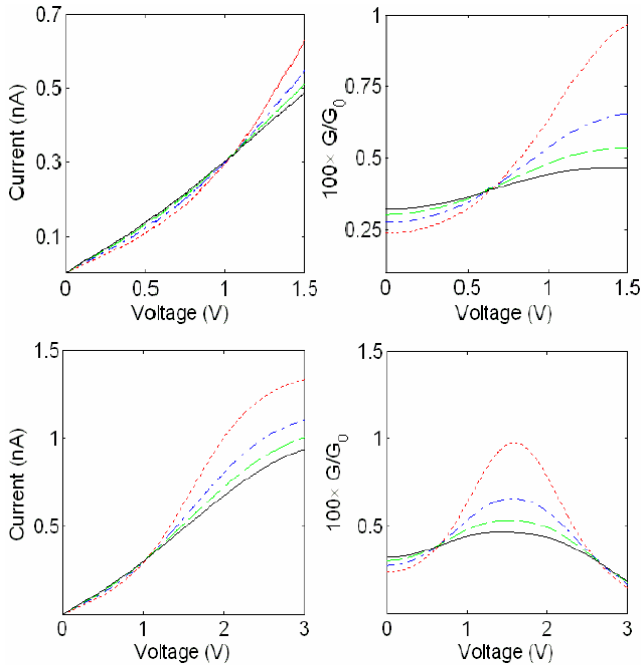}
\caption{ Current (left panels) and conductance (right panels) versus voltage.
The curves are plotted at $ T= 50K,\ T = 30 K,\ E_1 = 40 meV,\ 
\Delta_1 = \Delta_2= 4meV,\ n=1, \ \Delta V/V=0.005, $ 
assuming $J_0 = 9meV $ (solid line), $J_0 = 6meV $ (dashed line),
$ J_0= 3meV $ (dash-dotted line), and $J_0 = 1.5 meV $ (dotted line).
$G_0 = 2e^2/h$ is the quantum conductance.
}  
\label{rateI}
\end{center}
\end{figure}

Otherwise, the coherent tunneling between the metallic-like islands prevails in
the intergrain electron transport, and the influence of thermal phonon bath
is weak. Again, we may remark that $ J_0$ and $ T $ are combined as $kTJ_0 $ 
in the expression (21) for the dephasing
strength $\epsilon.$  Therefore an
increase in temperature at a fixed electron-phonon coupling strength enhances 
the incoherent contribution to the current in the same way as the previously
discussed increase in the electron-phonon coupling. Also,  at low
values of the applied voltage the electron-phonon coupling brings an enhancement
in both current and conductance, as shown in the top panels of the Fig. 3,
whereas the effect becomes reversed as the 
voltage grows above a certain value (see Fig. 3, the bottom panels). 
This happens because the phonon induced 
broadening of the intermediate energy level (the bridge)
 assists the electron transport at small bias voltage.
As the voltage rises, this effect is surpassed by the scattering effect of
phonons which resists the electron transport.

Now, we consider temperature dependencies of the electric current and 
conductance resulting from the intergrain electron tunneling via the intermediate
localized state. These dependencies are shown in the Fig. 4. The curves in the
figure are plotted at low bias voltage $(V= 0.3 V,\ \Delta V/V = 0.005)$ and
$T_0 = 50 K,$ so $e\Delta V < kT.$  This regime is chosen to compare the 
obtained temperature
dependencies with those typical for the phonon assisted hopping transport
discussed in the begining of the present work. We see that the tunneling current 
temperature dependence shown in the left panel of Fig. 4 crucially disagrees 
with the Mott's expression describing the ``hopping" electron transport. The
tunnel current decreases as temperatute rises being proportional to 
$(T_0/T)^\beta, $  and the exponent $\beta $ takes on values close to unity.

\begin{figure}[t]
\begin{center}
\includegraphics[width=4.7cm,height=8.8cm,angle=-90]{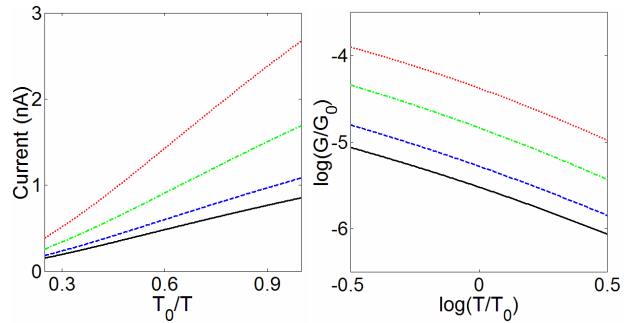}%%{pol5.eps}
\caption{ Temperature dependence of current (left panel) and conductance (right
panel) at low voltage bias $(V = 0.3 V).$ 
The curves are plotted at $ T_0 = 50 K, $ 
 $J_0 = 9meV $ (solid line), $J_0 = 6meV $ (dashed line),
$ J_0= 3meV $ (dash-dotted line), and $J_0 = 1.5 meV $ (dotted line). The  
values  of remaining parameters coincide with those used in the figure 3.
}  
\label{rateI}
\end{center}
\end{figure}

Already it was mentioned that the drop in the conductivity upon heating a 
sample was observed in polymers and carbon nanotubes. However, such 
metallic-like behavior could originate from various dc transport mechanisms.
Correspondingly, the specific features of temperature dependencies of 
the conductivity and/or current vary depending on the responsible conduction
mechanisms \cite{33}. The particular temperature dependence of the electron
tunneling current obtained in the present work and shown in the figure 3
differs from those occurring due to other transport mechanisms. Such
dependence was observed in the experiment on the electron transport in a 
single low-defect-content carbon nanotube rope whose metallic-like 
conductivity was manifested within a wide temperature range $(T\sim 35\div
300K),$ as reported by Fisher et al \cite{34}. The conductivity temperature
dependence observed in this work could be approximated as $\sigma (T)/
\sigma(300) \sim a+ bT_0/T $ where $A,b$ are dimensionless constants.
The approximation includes the temperature 
independent term which corresponds to the Drude conductivity. The second term
is inversely proportional to the temperature in agreement with present 
results for the current shown in the Fig. 3. It is also likely that a similar
approximation may be adopted to describe the experimental data obtained for
chlorate-doped polyacetylene samples at the temperatures below $100 K$ \cite{35}.
In both cases we may attribute the contribution proportional to $ 1/T$ to the
resonance electron tunneling transport mechanism.

Also, the conductance due to the electron intergrain tunneling reduces when 
the temperature increases, as shown in the right panel of the Fig. 4. 
Irrespective of the electron-phonon coupling strength we may approximate the 
conductance by the power law $G\sim T^\alpha $ where $\alpha $ takes on values
close to $ -1. $ This agrees with the results for the current.
 At higher bias voltage the temperature dependence of the current  changes, 
as shown in the Fig. 5. The curves shown in this figure could be approximated 
as $\ln (I/I_0) \sim c + d T_0/T;\ c,d$ being dimensionless constants.
This resembles typical temperature dependencies of the tunneling current 
in quasi-one-dimensional metals which
were predicted for conducting polymers being in a metal state
(see e.g. Ref. \cite{2}).

\begin{figure}[t]
\begin{center}
\includegraphics[width=6.3cm,height=6cm,angle=-90]{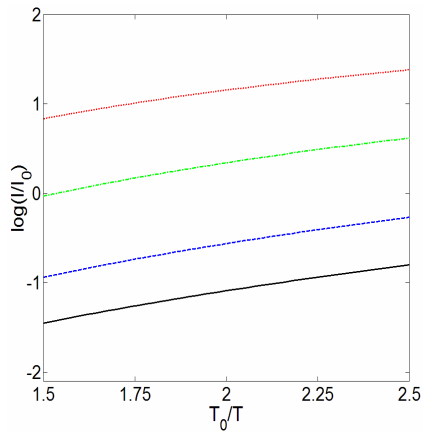}
\caption{ Temperature dependence of the electron current 
 at higher bias voltage  $(V = 2 V).$ 
The curves are plotted at $ T_0 = 50 K, $ 
 $J_0 = 9meV $ (solid line), $J_0 = 6meV $ (dashed line),
$ J_0= 3meV $ (dash-dotted line), and $J_0 = 1.5 meV $ (dotted line). The  values 
of remaining parameters are the same in the figures 3,4. The current $I_0 $
is computed using $J_0 = 3 meV;\ T_0/T = 1.5.$
}  
\label{rateI}
\end{center}
\end{figure}

Finally, there exist several mechanisms which could simultaneously
work providing the charge
transport in highly disodered and inhomogeneous materials as conducting polymers
and their relative significance could vary depending on both specific intrinsic
characteristics of particular materials (such as crystallization rate and
electron-electron and electron-phonon coupling strengths) and on the external
factors such as temperature. Various conduction mechanisms give rise to
various temperature dependencies of the electric current and conductance which
could be observed in polymer nanofibers/nanotubes. In the present work we
carried out theoretical studies to find out the character of temperature
dependencies of both current and conductance provided by specific transport
mechanism, namely, resonance tunneling of elecrons.

Accordingly, we 
treat a conducting polymer as a kind of granular metal, and we assume that
the intergrain conduction occurs due to the electron tunneling betweem the
metalliclike grains through the intermediate state.  To take into account the 
effect of temperature we represent the thermal environment (stochastic nuclear 
motions) as a phonon bath, and we introduce the coupling of the intermediate
site to the thermal phonons. In calculations of the electron transmission in 
the presence of dissipation originating from the thermal environment we follow
the way proposed in the work \cite{30}. The latter is based on the Buttiker
model within the scattering matrix formalism but instead of treating the
dephasing strength as a phenomenological parameter, the latter
 is expressed in terms of relevant energies, such as temperature and 
electron-phonon coupling strength.

%\vspace{2mm}

Using this approach we showed that the above described transport mechanism results 
in the temperature dependencies of transport characteristics which differ from
those obtained for other conduction mechanisms such as phonon-assisted hopping
between localised states. Being observed in experiments on realistic polymer 
nanofibers, the predicted dependencies would give grounds to suggest the electron
tunneling to predominate in the intergrain electron transport in these particular
nanofibers. We believe the present studies to contribute to 
better understanding of electron transport mechanisms in conducting polymers 
and carbon nanotubes. Also, we believe that the
model developed here may be used to theoretically analyze inelastic electron
transport through molecular junctions and nanodevices including  quantum dots
coupled to the electron reservoirs.

\section{Acknowledgments:} Author  thanks  G. M. Zimbovsky for help with the 
manuscript.
This work was supported  by DoD grant W911NF-06-1-0519 and NSF-DMR-PREM 0353730.

%\begin{widetext}  \end{widetext}


\begin{thebibliography}{99}
\bibitem{33} The matter is thoroughly discussed by Kaiser (see \cite{2} and
references therein).
\end{thebibliography}
\end{document}